\documentclass[twocolumn,pra,superscriptaddress]{revtex4-2}
\usepackage{amsfonts}
\usepackage{amssymb}
\usepackage{amsmath}
\usepackage{epsfig}
\usepackage{color}
\usepackage{graphics, graphicx}
\usepackage{bbold}
\usepackage{psfrag}
\usepackage{mathcomp}
\usepackage{subfigure}
\usepackage{verbatim}
\usepackage{float}
\usepackage{graphicx}
\usepackage[colorlinks,citecolor=blue]{hyperref}
\makeatletter

\newcommand{\Rmnum}[1]{\expandafter\@slowromancap\romannumeral #1@}
\makeatother
\begin{document}
	
	\title{Observation of gauge field induced non-Hermitian helical skin effects}
	
	\author{Yu-Hong Han}
	\altaffiliation{These authors contributed equally to this work.} 
	\affiliation{State Key Laboratory of Quantum Optics Technologies and Devices, Institute
		of Laser Spectroscopy, Shanxi University, Taiyuan, Shanxi 030006, China}
		
	\author{Yi Li}
	\altaffiliation{These authors contributed equally to this work.} 
	\affiliation{State Key Laboratory of Quantum Optics Technologies and Devices, Institute
		of Laser Spectroscopy, Shanxi University, Taiyuan, Shanxi 030006, China}
	
	\author{Jia-Hui Zhang}
	\affiliation{State Key Laboratory of Quantum Optics Technologies and Devices, Institute of Laser Spectroscopy, Shanxi University, Taiyuan, Shanxi 030006, China}

    \author{Yang Kou}
	\affiliation{State Key Laboratory of Quantum Optics Technologies and Devices, Institute of Laser Spectroscopy, Shanxi University, Taiyuan, Shanxi 030006, China}

\author{Liantuan Xiao}
	\affiliation{State Key Laboratory of Quantum Optics Technologies and Devices, Institute
		of Laser Spectroscopy, Shanxi University, Taiyuan, Shanxi 030006, China}
	\affiliation{Collaborative Innovation Center of Extreme Optics, Shanxi
		University, Taiyuan, Shanxi 030006, China}

\author{Suotang Jia}
	\affiliation{State Key Laboratory of Quantum Optics Technologies and Devices, Institute
		of Laser Spectroscopy, Shanxi University, Taiyuan, Shanxi 030006, China}
	\affiliation{Collaborative Innovation Center of Extreme Optics, Shanxi
		University, Taiyuan, Shanxi 030006, China}

\author{Linhu Li}
      \email{lilinhu@quantumsc.cn}
	\affiliation{Quantum Science Center of Guangdong-Hong Kong-Macao Greater Bay Area (Guangdong), Shenzhen, China}

     \author{Feng Mei}
      \email{meifeng@sxu.edu.cn}
	\affiliation{State Key Laboratory of Quantum Optics Technologies and Devices, Institute
		of Laser Spectroscopy, Shanxi University, Taiyuan, Shanxi 030006, China}
	\affiliation{Collaborative Innovation Center of Extreme Optics, Shanxi
		University, Taiyuan, Shanxi 030006, China}

\begin{abstract}
Synthetic gauge fields and non-Hermitian skin effects are pivotal to topological phases and non-Hermitian physics, each recently attracting great interest across diverse research fields. Realizing skin effects typically require nonreciprocal couplings or on-site gain and loss. Here, we theoretically and experimentally report that, under gauge fields, reciprocal dissipative couplings can nontrivially give rise to an unprecedented nonreciprocal skin effect, hosting pseudospin degree of freedom and featuring helical transport, dubbed as the ``helical pseudospin skin effect". Before introducing the gauge fields, this model exhibits localized pseudospin edge modes and extended bulk modes, without skin effects. As the gauge field strength is applied from $0$ to $\pi$, we observe the emergence of two distinct pseudospin skin effects and their topological transitions: the hybrid-order and second-order helical pseudospin skin effects. Our findings not only highlight gauge field enriched non-Hermitian topology, but also brings pseudospin-momentum locking into skin effects.
\end{abstract}
	\maketitle

\textit{Introduction}---Non-Hermitian physics has recently attracted significant attention by introducing behaviors that are not possible in conservative Hermitian systems~\cite{LiG2017,Demetrios2018}. In particular, the non-Hermitian skin effect (SE) is especially intriguing, as it causes wave modes to localize at the boundaries rather than being distributed throughout the bulk~\cite{Wangzhong2018,GongZP2018,LLH2019,LeeCH2019,Flore2021_review,LuMH2022,LLH2023_review,FanSH2024_review,LLH2024_revirew,LiCA2023,RoyB2023}. This effect holds profound implications for the development of innovative wave-based technologies, including wave morphing~\cite{Alexander2020,Christensen2021,Maguancong2022}, topological lasers~\cite{ChongYD2022} and advanced sensors~\cite{ZhangXD2023}. Consequently, its experimental realization has been actively pursued across a diverse range of wave systems, including photonic~\cite{DengWY2023,Fengliang2023,LiuZY2024,LiuZY2025review,YangZJ2024,LiTao2024,Edo2024,ZhangBL2024}, acoustic~\cite{ZhangBL2021nc,ChenYF2021,ChongYD2022prb,ZhuJie2022,KunDing2023,LiuZY2023,YangZJ2023,QiuCY2024,JJH2024,QiuCY2024prb,LiuZY2025,LiuZY2025-3}, circuit~\cite{Thomale2020,Neupert2020,ZhangXD2021,LiuZY2022,Bahl2023,ZhangXD2023prb,LLH2024,HuHP2024}, transmission line~\cite{ChanCT2024} and quantum systems~\cite{XueP2020,LLH2020,YanB2022,XueP2023,Gyu2025,LeeCH2025}. Notably, implementing non-Hermitian SEs usually requires unequal left-right coupling strengths or on-site gain and loss~\cite{LiuYC2022,LiuYC2024,LiuZY2025-2}, which directly or equivalently creates an asymmetry in wave propagation and results in the accumulation of wave modes at one boundary.

Meanwhile, there has been growing interest in recent years in the realization of both Abelian~\cite{Goldman2018_review,YuanLQ2016,YuanLQ2018,LitaoNRP} and non-Abelian synthetic gauge fields~\cite{YanB2024,FanSH2025}. This artificial field mimics the behavior of real gauge fields on charge-neutral classical waves and can lead to novel topological physics, such as Brillouin Klein bottle~\cite{ZhaoYX2022}, chiral zero modes~\cite{ZhangS2023} and Stiefel-Whitney topological phases~\cite{ZhangBL2023}. On the application side, such fields can also open a new dimension for controlling waves, such as gauge field induced Landau rainbows~\cite{ZhangS2024}, negative refraction~\cite{FanSH2013prl,ZhangBL2021}, robust transport~\cite{FanSH2014,Hafezi2014,FanSH2015}.

\begin{figure*}[htbp]
	\centering
	\includegraphics[scale=0.9]{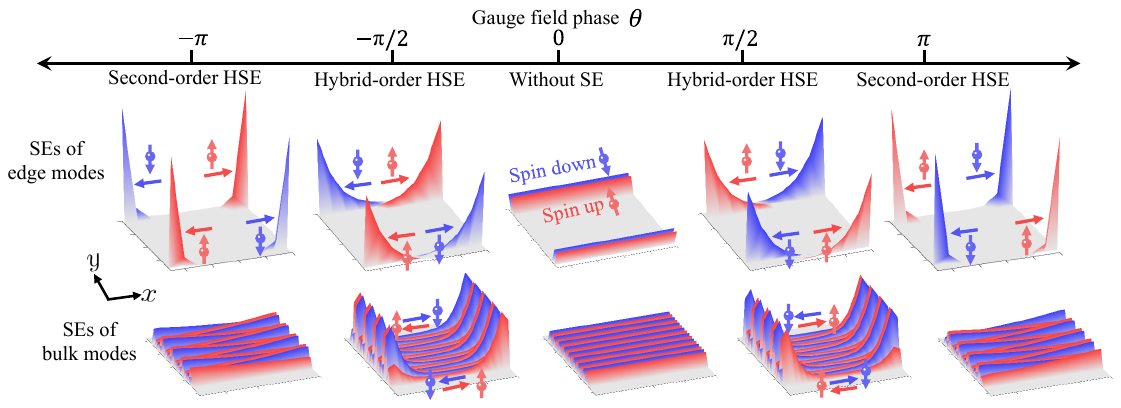}
	\caption{Schematic of HSEs induced and enriched by gauge fields. In a non-Hermitian system with reciprocal couplings and without on-site gain and loss, nonreciprocal SEs are generally not expected. This work shows that the introduction of gauge fields into such system can induce  nonreciprocal HSEs. For the zero gauge field ($\theta = 0$), the system hosts localized pseudospin edge modes and extended bulk modes, while no SEs are observed. When the gauge field is turned on ($\theta > 0$), the pseudospin-up and pseudospin-down eigenmodes are respectively accumulated towards opposite corners or edges of the $x$-direction, regardless of whether they are edge or bulk modes, resulting in hybrid-order HSEs. When $\theta=\pi$, the pseudospin-up and pseudospin-down edge modes continue to collapse towards opposite corners, while the bulk modes begin to delocalize, leading to second-order HSEs. Reversing the sign of the gauge field phase can flip the directions of the HSEs.}
	\label{Fig1}
\end{figure*}

In this work, we theoretically and experimentally report a mechanism in which the combination of synthetic gauge fields with reciprocal dissipative couplings can nontrivially induce nonreciprocal skin localization. Interestingly, this mechanism gives rise to a new class of two-dimensional SEs that exhibit distinct characteristics: hosting pseudospin degree of freedom and featuring helical skin localization tuned by the gauge field phase (as illustrated in Fig.~\ref{Fig1}), termed the helical pseudospin SE (HSE). Experimentally, we implement both the synthetic gauge fields and the reciprocal dissipative couplings in circuit metamaterials. Before applying gauge fields, this model exhibits localized pseudospin edge modes and extended bulk modes, without SEs. As the gauge field is introduced and its phase increases from $0$ to $\pi$, we experimentally observe the emergence of two distinct HSEs, undergoing topological transitions between hybrid-order and second-order HSEs. Besides, reversing the sign of the gauge field phase can flip the helical directions of the pseudospin-up and pseudospin-down skin modes. This SE is analogous to the helical transport of pseudospin edge modes in time-reversal-invariant topological insulators, increasing the gauge field $\theta$ induces helical pseudospin-dependent transport of skin modes, which we refer to as HSEs.

\begin{figure*}[htbp]
	\centering
	\includegraphics[scale=0.85]{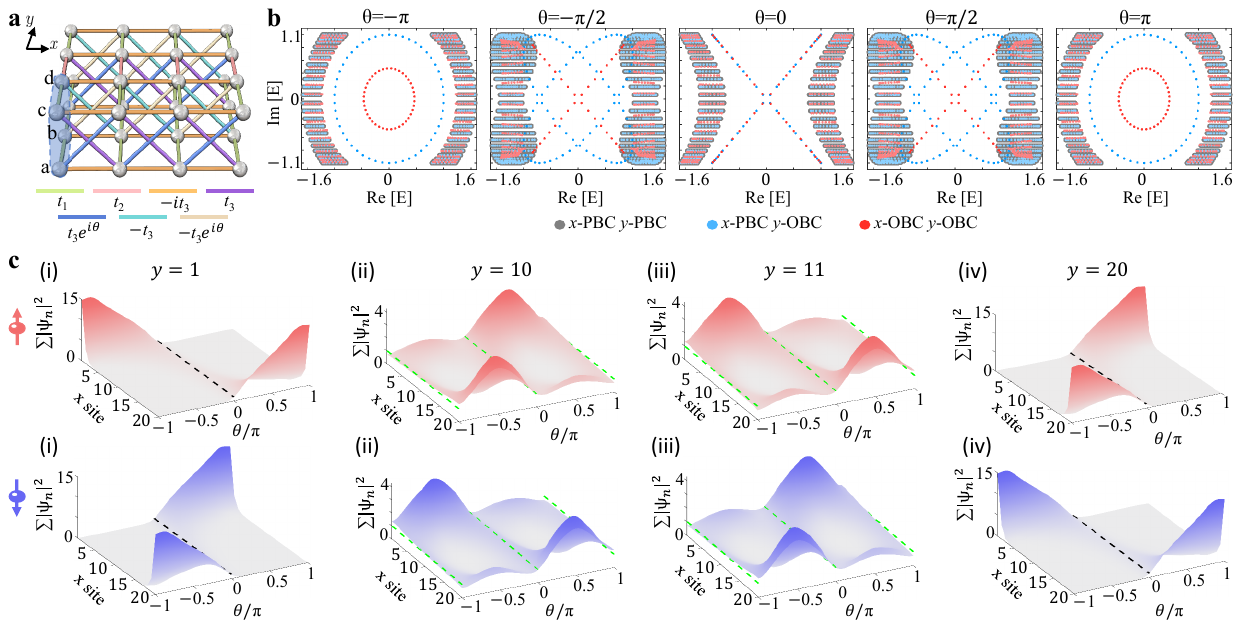}
	\caption{Model and helical SEs. (a) Non-Hermitian lattice model with reciprocal dissipative couplings $-it_3$ and synthetic gauge fields $e^{i\theta}$. (b) Energy spectra under different boundary conditions, varying with five distinct gauge field phases. (c) Density distributions for all the pseudospin-up and pseudospin-down modes as a function of $\theta$ for four different y. The black and green dashed lines indicate the points where the SE vanishes for the edge and bulk modes, respectively. As shown, with increasing $\theta$, the pseudospin-up and pseudospin-down skin modes feature helical evolutions along the x-direction. The parameters are $t_1 = 0.22t_2$, $t_3 = 0.5t_2$, $N_x = 20$ and $N_y = 10$.}
	\label{Fig2}
\end{figure*}

\textit{Gauge field induced HSEs}---To be concrete, we consider a two-dimensional bilayer lattice model, as depicted in Fig. \ref{Fig2}(a), where each unit cell comprises four sites. The momentum-space Hamiltonian can be written as
\begin{equation}
 \begin{aligned}
  H= & -2it_3\cos k_x \sigma_0 \tau_0 + (t_1 + t_2 \cos k_y) \sigma_0 \tau_x \\ & + t_2 \sin k_y \sigma_0 \tau_y + t_3 \left(\sin k_x - \sin (k_x + \theta)\right) \sigma_y \tau_z \\ & + t_3 \left(\cos k_x + \cos (k_x + \theta)\right) \sigma_x \tau_z,
 \end{aligned}
\end{equation}
where $\tau_{x,y,z}$ and $\sigma_{x,y,z}$ represent the Pauli matrices acting on the sublattice space $(a,b)$ or $(c,d)$, and the upper and lower layer degrees of freedom, respectively. $t_1$ represents the couplings within unit cells, $t_2$ represents the couplings between unit cells, $-it_3$ is the reciprocal dissipative couplings that introduce non-Hermiticity and feature equal left-right coupling strengths, and $\theta$ is the gauge field phase applied on the couplings between layers. Despite the presence of gauge fields, we find that the non-Hermitian bilayer system respects a gauge-field dependent mirror symmetry, expressed as $\mathcal{M}_zH(k_x,k_y)\mathcal{M}^{-1}_z=H(k_x,k_y)$, where the mirror symmetry operator is given by $\mathcal{M}_z = G\sigma_x\tau_0$, with $\mathcal{M}^2_z=1$, $G = \cos\frac{\theta}{2}\sigma_0\tau_0+i\sin\frac{\theta}{2}\sigma_z\tau_0$ being the gauge-field dependent symmetry transformation.

Since $[\mathcal{M}_z, H] = 0$, the Bloch Hamiltonian $H$ can be block diagonalized in the eigenspace of $\mathcal{M}_z$. We refer to the eigenvector spaces of $\mathcal{M}_z$ as pseudospin spaces, with eigenvalues $\pm 1$ corresponding to the pseudospin-up and pseudospin-down subspaces, respectively. Within these two subspaces, the Bloch Hamiltonian block-diagonalizes into $H_+ \oplus H_-$, where $+$ and $-$ denote the pseudospin-up and pseudospin-down sectors. Accordingly, the pseudospin-up and pseudospin-down labels in Figs. (2-4) correspond to the results obtained within these respective pseudospin subspaces. The block Hamiltonian takes the following form
\begin{equation}
\begin{aligned}
 H_{\pm}(k) = & -2it_3 \cos k_x \tau_0 + (t_1 + t_2 \cos k_y) \tau_x \\ & + t_2 \sin k_y \tau_y \pm 2t_3 \cos\left(k_x + \frac{\theta}{2}\right) \tau_z.
 \end{aligned}
 \end{equation}
It turns out that $H_{\pm}$ have a periodicity of $4\pi$ on the gauge field phases, i.e., $H_{\pm}(k_x,k_y,\theta)=H_{\pm}(k_x,k_y,\theta+4\pi)$. The pseudospin-up and pseudospin-down subspace Hamiltonians are related to each other by $H_{\pm}(k_x,k_y,\theta)=H_{\mp}(k_x,k_y,\theta+2\pi)$. As revealed below, this feature ensures that the pseudospin-up and pseudospin-down SEs exhibit a pseudospin-rotational symmetry under a gauge field phase shift of $2\pi$. In Supplementary Materials, we also show that the absence and emergence of SEs at the gauge field phase $\theta=0,\pm\pi$ can be attributed to the time reversal symmetry properties of the Bloch Hamiltonian~\cite{ZhaoYX2023}. It should be noted that our use of mirror symmetry serves primarily as a clear and intuitive framework to help readers understand the HSEs. Such effect itself is a general and robust phenomenon that does not rely strictly on mirror symmetry. As demonstrated in Supplementary Materials, the HSE persists even when mirror symmetry
is broken and can also appear under other symmetries, such as rotational symmetry.

The emergence of SEs can be characterized by geometric features of energy spectra for bulk and edge modes \cite{LLH2019,LLH2024pra,FangC2022}. In Fig. \ref{Fig2}(b), we study the energy spectra under three types of boundary conditions with varying gauge field phases. As shown, under case (i), the spectra consist solely of two gapped bulk bands, corresponding to eigenenergies with positive and negative $\mathrm{Re}[E]$. In contrast, for cases (ii) and (iii) under OBC, alongside the two bulk bands, additional in-gap edge bands appear in the middle of the spectra. As discussed below, the spectral distinction between cases (ii) and (iii) allows us to distinguish the SEs associated with edge modes from those of bulk modes.

Specifically, at $\theta = \pm \pi$, under cases (ii) and (iii), the geometries of bulk bands remain identical, whereas the edge bands shrink from a larger to a smaller circle. This behavior reveals that the edge modes exhibit SE, while the bulk modes do not. As depicted in Fig. \ref{Fig2}(c), all the edge modes correspondingly are maximally localized at the corners of the two-dimensional lattice, giving rise to a second-order SE~\cite{KenS2020}. When the gauge field phase is tuned to $\theta = \pm \pi/2$, the geometries of the bulk and edge bands both become distinct under different boundary conditions, signifying the coexistence of SEs in both. In this case, the SE of edge modes persists at the corners, corresponding to the second-order SE, while the SE of bulk modes appears along the edges (see Fig.~\ref{Fig2}(c)), leading to the conventional first-order SE. Together, they constitute a hybrid-order SE emerged in the system. In contrast, at $\theta = 0$, the whole spectra are identical for cases (ii) to (iii), indicating the absence of any SE. 

To uncover the spin-dependent helical features of the SEs, Fig.~\ref{Fig2}(c) shows the density distribution $\sum|\psi_n|^2$ of both edge and bulk eigenmodes as a function of the gauge-field phase $\theta$, evaluated at the lower edge ($y=1$), bulk sites ($y=10, 11$), and the upper edge ($y=20$). Here $\psi_n$ denotes the eigenmodes in different pseudospin subspaces. For pseudospin-up modes on the lower edge, a left SE appears at $\theta=-\pi$, gradually weakens as $\theta$ increases, and vanishes at the critical point $\theta=0$. Beyond this, a right SE emerges and reaches its maximum at $\theta=\pi$. The upper edge exhibits the opposite evolution: the SE starts on the right, disappears at $\theta=0$, and reappears on the left at $\theta=\pi$. (see  Supplemental Materials for more details). The bulk modes show a richer pattern. Their SE direction depends on the $y$ coordinate: for even $y$ (e.g., $y=10$), the SE switches from right-moving to left-moving as $\theta$ crosses $0$, whereas odd $y$ values display the reversed trend. For pseudospin-down modes, the SE intensity follows the same phase dependence, but the propagation direction is reversed relative to pseudospin-up. These observations confirm that the SEs exhibit gauge-field-induced helical evolution.

\begin{figure*}[htbp]
	\centering
	\includegraphics[scale=1.1]{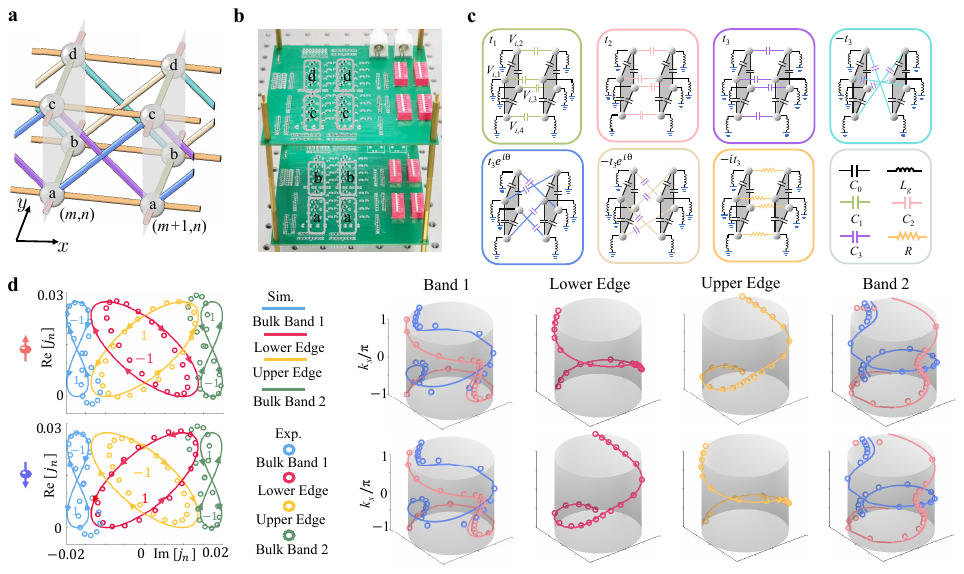}
	\caption{Implementation of synthetic gauge fields and measurement of topological winding numbers. (a) Schematic of the couplings between two unit cells. The coordinates for each unit cell are labelled as $(m,n)$. (b) The corresponding printed circuit board. (c) Specific circuit connections for implementing $t_1$, $t_2$, $\pm t_3$, $\pm t_3e^{i\theta}$ and $-it_3$ to implement the gauge field phase $\theta=\pi/2$. (d) The left panel shows measured (circles) and simulated (lines) energy spectra under x-PBC and y-OBC at $\theta = \pi/2$, while the right panel presents the corresponding three-dimensional illustration of the spectral winding topology}.
	\label{Fig3}
\end{figure*}

The topology of the HSEs of edge and bulk modes can be characterized by the following topological winding number~\cite{Wangzhong2018}: $W = \frac{1}{2\pi i} \oint_{-\pi}^{\pi}  \frac{\partial}{\partial k_x} \log \left[E(k_x) - E_r\right]{\rm{d}}k_x$, where $E(k_x)$ represents the energy spectra under x-PBC and y-OBC, and $E_r$ is a reference energy. For any reference energy $E_r$ located inside the closed loop of eigenenergies, the winding number takes nontrivial values of $W\neq 0$. This winding number characterizes the direction of the exponential decay of the eigenmode wavefunctions associated with the loop and thus determines the direction of the SEs. Take $\theta = 0.5\pi$ as an example, the energy spectrum under x-PBC and y-OBC forms closed loops in both the spin-up and spin-down subspaces, as shown in Fig. \ref{Fig3}(d), respectively. As illustrated at the right side of Fig. \ref{Fig3}(d), the sign of winding numbers are determined by the winding directions as a function of $k_x$. For each pseudospin, the two bulk bands (blue and green) simultaneously exhibit winding numbers $W=\pm1$, giving rise to left and right SEs of bulk modes. In contrast, for pseudospin-up edge bands, the lower edge band (red) exhibits $W = -1$ (right SE of edge modes), while the upper band (yellow) exhibits $W = 1$ (left SE of edge modes). For pseudospin-down edge bands, the situation is reversed: the lower band exhibits a left SE ($W = 1$), and the upper band exhibits a right SE ($W = -1$). The winding behavior is clearly observed despite experimental imperfections.

\begin{figure*}[htbp]
	\centering
	\includegraphics[scale=0.8]{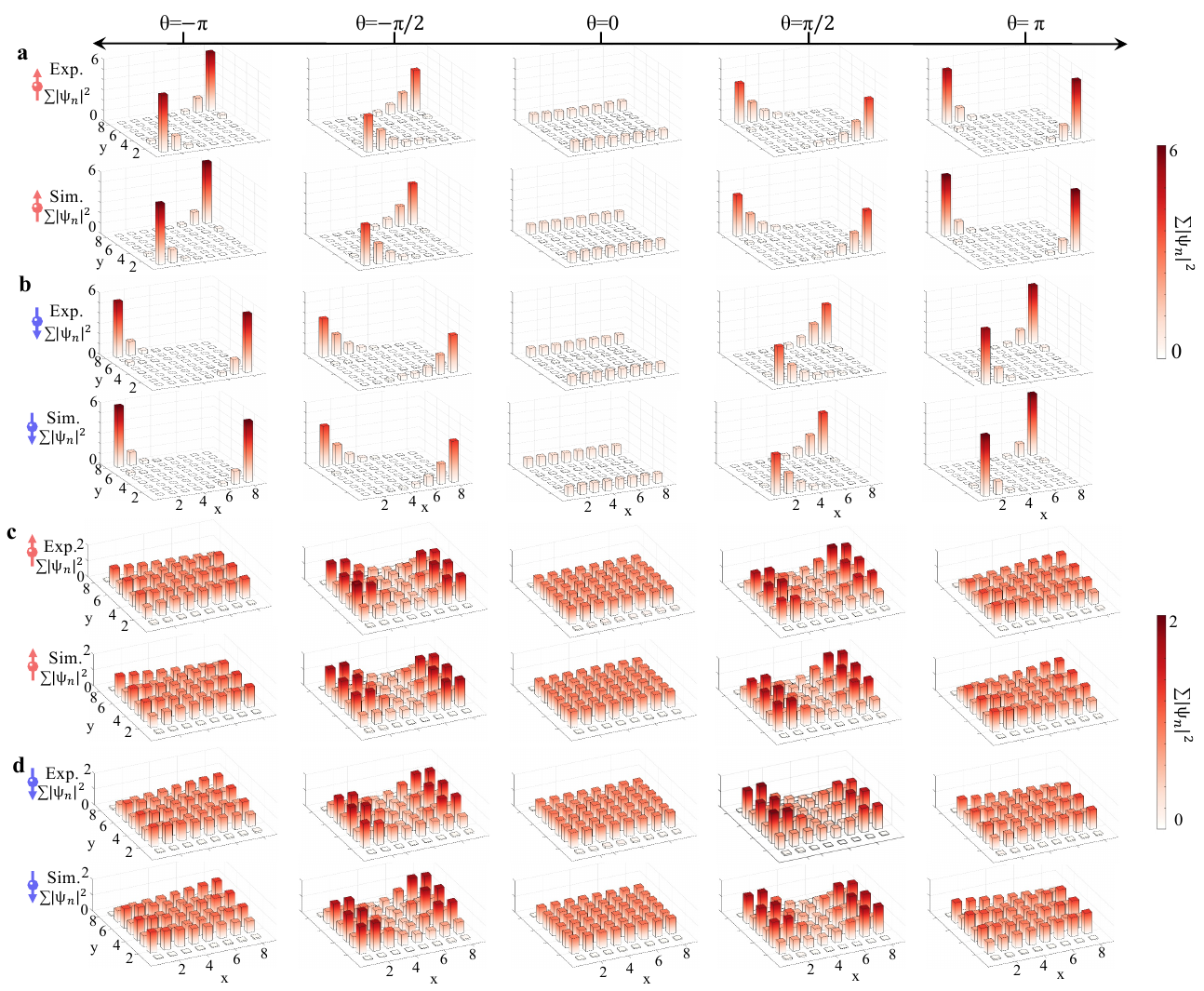}
	\caption{Observation of non-Hermitian HSEs. Measured and simulated density distributions for pseudospin-up $(a)$ and pseudospin-down $(b)$ edge modes varying with the gauge field phases $\theta = -\pi, -\pi/2, 0, \pi/2, \pi$. (c,d) Same as (a,b) but for the bulk modes. As shown, for the gauge field phase $\theta = 0$, the system exhibits no SE. While for $\theta=\pm\pi/2$, the system exhibits a hybrid-order HSE, where pseudospin-up and pseudospin-down edge and bulk modes are maximally localized at opposite boundaries. At $\theta =\pm\pi$, the system transitions to a second-order HSE, with only edge modes remaining localized.}
	\label{Fig4}
\end{figure*}

\textit{Experimental implementation}---Next, we implement the lattice model under various gauge-field phases using circuit metamaterials~\cite{LeeCH2018,Thomale2018,YanP2024_review,ChenT2024,LeeCH2025_review}. Here we focus on the realization of the $\theta=\pi/2$ configuration using the basic two–unit-cell building block shown in Fig.~\ref{Fig3}(a), while the circuit designs for other phases ($\theta = 0, \pm\pi/3, -\pi/2, \pm 2\pi/3, \pm\pi$) are provided in Supplemental Materials. The fabricated PCB and the mapping between the circuit network and the lattice couplings are presented in Figs.~\ref{Fig3}(b) and~\ref{Fig3}(c). To encode the complex hopping amplitudes required by the gauge field, each lattice site is represented by four circuit nodes coupled through capacitors and grounded through an inductance $L_g$. The intra- and inter-cell couplings $t_1$, $t_2$, and $t_3$ are implemented by capacitors $C_1$, $C_2$, and $C_3$, while the phase-dependent hoppings ($-t_3$, $\pm t_3 e^{i\pi/2}$) are generated through cross-connections between the corresponding nodes. Reciprocal dissipative couplings proportional to $-i t_3$ are realized by resistive links. To validate the circuit design, we derive the eigen-equation from Kirchhoff's law. By driving the quadrangle with the composite excitation 
$\mathbf{V}_i = V_{i,1} + e^{-i\pi/2} V_{i,2} + e^{i\pi} V_{i,3} + e^{i\pi/2} V_{i,4}$,
the response satisfies 
$\mathbf{I}_\alpha(\omega) = \sum_{\beta} J_{\alpha\beta}(\omega)\mathbf{V}_\beta(\omega)$,
where $J(\omega)$ is the circuit Laplacian. At the resonant frequency
$\omega_0 = 1/\sqrt{(C_1 + C_2 + 2C_3 + 2C_0)L_g}$,
the Laplacian $J(\omega_0)$ reproduces the non-Hermitian Hamiltonian of Fig.~\ref{Fig2}(a) for $\theta = \pi/2$ (see Supplemental Materials for more details).

\color{black}
    
\textit{Observation of HSEs}---To further investigate the pseudospin SE and its helical transport behavior as the gauge field varies, the circuit metamaterial is configured with x-OBC and y-OBC, consisting of $8 \times 8$ nodes. By diagonalizing the measured impedance matrix $G$ to obtain the eigenvectors $\left|\psi_n\right\rangle$, we track the density distributions of edge and bulk wavefunctions as a function of the gauge field phase $\theta$. Specifically, the measured and simulated results for the edge modes at the gauge field phases $\theta=0,\pm\pi/2,\pm\pi$ are shown in Figs. \ref{Fig4}(a,b), while those for the bulk modes are shown in Figs. \ref{Fig4}(c,d). The results for other gauge field phases $\theta=\pm\pi/3,\pm2\pi/3$ are provided in Supplemental Materials.

As shown, for the pseudospin-up edge modes at $\theta = -\pi$, the lower edge features a left SE, while the upper edge displays a right SE. As the gauge field phase $\theta$ is increased to $-\pi/2$, despite the directions of the SEs at both the lower and upper edges remain unchanged, their strengths are suppressed. As increased to $\theta = 0$, the left and right SEs completely vanish, and the edge modes become extensively distributed, indicating a phase where the system lacks any SE of edge modes. As the flux increases further to $\theta = \pi/2$, the direction of SEs reverses as compared to the one at $\theta=-\pi/2$, i.e., the lower edge now shows a right SE, and the upper edge exhibits a left SE, implying that the edge mode localization has been flipped to the opposite side of the system. At $\theta = \pi$, this reversed localization becomes stronger as compared to $\theta = \pi/2$. For the pseudospin-down edge modes, the overall behavior is similar to that of the pseudospin-up modes, but with the opposite direction of the SE at each stage. For example, when $\theta = -\pi$, the upper edge exhibits a left SE, and the lower edge shows a right SE. This reversal of direction of SE for pseudospin-down modes, as compared to pseudospin-up modes, demonstrates a clear helical pseudospin-dependent feature. Our experimental data agree well with the simulated results in the lower panel of Figs. \ref{Fig4}(a,b), including the directions and strengths of the SEs at the upper and lower edges.

For the pseudospin-up bulk modes at $\theta = -\pi$, there is no SE observed; the bulk modes are evenly distributed across the system, indicating no SE of bulk modes. As the gauge field increases to $\theta = -\pi/2$, an intriguing behavior emerges: the bulk modes show a left SE for odd $y$ coordinates and a right SE for even $y$ coordinates, indicating a pattern of alternating localization across the lattice. At $\theta = 0$, the SE once again vanishes, and the bulk modes return to a state of even distribution. As the flux increases further to $\theta = \pi/2$, the pattern of localization reappears, but with the directions reversed: now, the odd $y$ coordinates exhibit a right SE, and the even $y$ coordinates show a left SE. Finally, at $\theta = \pi$, the SE disappears once more, returning the bulk modes to an evenly distributed configuration, similar to what was observed at $\theta = 0$ and $\theta = -\pi$. For the pseudospin-down bulk modes, the behavior is similar to that of the pseudospin-up bulk modes, but with the opposite direction of the SE. Our experimental data are also in good agreement with the simulated results presented in the lower panel of Figs. \ref{Fig4}(c,d). 

The combination of experimental results for the SEs of bulk and edge modes unequivocally demonstrates that, as the gauge field phase $\theta \in(0, \pi)$, a hybrid-order HSE emerges, characterized by both HSEs of edge and bulk modes. As $\theta=\pi$, only HSEs of edge modes remain, with the HSEs of bulk modes disappearing, giving rise to second-order HSEs. Furthermore, the direction of the SEs induced by the gauge field helically depends on the pseudospin degree of freedom, similar to the helical pseudospin edge states found in time-reversal invariant topological insulator phases, leading to the emergence of HSEs. Moreover, the helical direction can be flipped by reversing the sign of the gauge field phase. In the Supplementary Materials, we further exhibit such HSE can be harnessed to realize pseudospin dependent nonreciprocality and used for designing one-wave spintronics.

\textit{Discussion}---In summary, we have theoretically and experimentally demonstrated an unconventional mechanism based on synthetic gauge fields that induces non-Hermitian SEs without relying on nonreciprocal couplings or on-site gain and loss. Furthermore, we have uncovered that this mechanism gives rise to a new class of two-dimensional SEs that host pseudospin degree of freedom and exhibit helical transport, thereby merging non-Hermitian topology with spin-helical physics. These findings could be of broad interest to the photonics, condensed matter, and metamaterials communities. Looking forward, given the recent realization of synthetic Abelian and non-Abelian gauge fields across a wide range of classical and quantum platforms~\cite{Goldman2018_review,YanB2024,LitaoNRP,FanSH2025}, our study provides timely insights and promising avenues for the further study of gauge field enriched non-Hermitian topological phases and SEs, and opens pathways for potential technological applications, such as gauge field induced and engineered spin-selective helical one-way quantum transport.
%\section*{Code availability}
%The codes are available upon reasonable request from the corresponding author.

\textit{Acknowledgment}---We acknowledge Profs. Jing Zhang, Z.D. Wang, Yu-Xin Zhao, Luqi Yuan,  Zhaoju Yang and Wenbin Rui for helpful discussions and positive comments. This work is supported by the National Key Research and Development Program of China (Grant No. 2022YFA1404201), National Natural Science Foundation of China (NSFC) (Grant No. 12474361, No. 12034012, No. 12074234), Changjiang Scholars and Innovative Research Team in University of Ministry of Education of China (PCSIRT)($IRT\_17R70$), Fund for Shanxi 1331 Project Key Subjects Construction, 111 Project (D18001) and Fundamental Research Program of Shanxi Province (Grant No. 202303021223005). Y.L. acknowledges support from the China Association for Science and Technology (CAST), China.

\noindent \textbf{REFERENCES}
\bibliographystyle{naturemag}
\bibliography{ref.bib}	
\end{document}